\documentclass[%
superscriptaddress,
 amsmath,amssymb,
 aps, pra, twocolumn,
 floatfix
]{revtex4-2}

\usepackage{graphicx}
\usepackage{dcolumn}
\usepackage{bm}
\usepackage{hyperref}
\usepackage{physics}
\usepackage{braket}
\usepackage{placeins}
\usepackage{xurl}
\usepackage{amsmath}
\usepackage{xcolor}
\begin{document}

\title{Fast and high-fidelity transfer of edge states via dynamical control of topological phases and effects of dissipation}

\author{Yuuki Kanda}
        \affiliation{Department of Applied Physics, Hokkaido University, Sapporo 060-8628, Japan.}
\author{Yusuke Fujisawa}
        \affiliation{Department of Applied Physics, Hokkaido University, Sapporo 060-8628, Japan.}
\author{Kousuke Yakubo}
        \affiliation{Department of Applied Physics, Hokkaido University, Sapporo 060-8628, Japan.}
        \affiliation{National Institute of Technology, Asahikawa College, Asahikawa 071-8142, Japan}
\author{Norio Kawakami}
        \affiliation{Fundamental Quantum Science Program, TRIP Headquarters, RIKEN, Wako 351-0198, Japan}
	\affiliation{Department of Physics, Ritsumeikan University, Kusatsu, Shiga 525-8577, Japan}
	\affiliation{Department of Materials Engineering Science, Osaka University, Toyonaka, Osaka 560-8531, Japan}
\author{Hideaki Obuse}
        \affiliation{Department of Applied Physics, Hokkaido University, Sapporo 060-8628, Japan.}
        \affiliation{Institute for Frontier Education and Research on Semiconductors, Hokkaido University, Sapporo 060-0808, Japan.}
        \affiliation{Institute of Industrial Science, The University of Tokyo, 5-1-5 Kashiwanoha, Kashiwa, Chiba 277-8574, Japan}
        \affiliation{The Institute for Solid State Physics, The University of Tokyo, Kashiwa, Chiba 277-8581, Japan}

\date{\today}

\begin{abstract}

Topological edge states are robust against symmetry-preserving perturbations and noise, making them promising for quantum information and computation, particularly in topological quantum computation through the braiding operations of Majorana quasiparticles. Realizing these applications requires fast and high-fidelity dynamic control of edge states. In this work, we theoretically propose a high-fidelity protocol for transferring topological edge states by dynamically moving a domain wall between two regions with different topological numbers in one dimension. This protocol fundamentally relies on Lorentz invariance and relativistic effects, because moving the domain wall at a constant speed is described by a mass term with the uniform linear motion in the Dirac equation. We demonstrate the effectiveness of our protocol in transferring edge states with high fidelity using a one-dimensional quantum walk with two internal states, which is feasible with current experimental technology. We also investigate how bit-flip and dephasing dissipation to the environment affect transfer efficiency. Remarkably, bit (dephasing) dissipation does not affect the fidelity at the slow (fast) transfer limit, which can be explained by the relativistic effects on the edge states.
\end{abstract}

\maketitle


\section{\label{sec:level1}Introduction}

The realization of large-scale quantum devices, such as quantum computers, requires the implementation of quantum information processing, which in turn necessitates robust and high-fidelity quantum state transfer. However, 
quantum states are inherently fragile to decoherence. 
To overcome this vulnerability, various methods have been proposed in various platforms, such as, coupling between atoms and photons\cite{PhysRevLett.78.3221, science.1103346}, spin systems \cite{PhysRevLett.91.207901, PhysRevLett.92.187902, 10.1063/1.2008129}, and so on\cite{andp.201400116, PhysRevLett.118.133601, PhysRevX.7.011035}. 

Among these strategies, quantum state transfer utilizing edge states appearing in topological matter is one of the most promising approaches~\cite{Yao2013, Dlaska_2017, Lang2017, PhysRevA.98.012331}. This is because the edge states are topologically protected and then robust against symmetry-preserving perturbations. These edge states appear as localized states near domain walls which are spatial boundaries where a discrete topological number varies in the position space~\cite{PhysRevLett.95.146802, PhysRevLett.95.226801, RevModPhys.82.3045, RevModPhys.83.1057,  RevModPhys.89.041004}. Thus, the dynamical control of domain walls could provide alternative route to quantum state transfer through thetransport of the edge states following the moving domain walls.

Furthermore, 
edge states are also essential for the realization of topological quantum computation, a fault-tolerant quantum computation utilizing Majorana quasiparticles~\cite{Kitaev2001, Kitaev:2009war}. Majorana quasiparticles are edge states that emerge in systems such as 
topological superconductors realized in quantum nanowires~\cite{PhysRevLett.86.268, Alicea2011, PhysRevLett.100.096407, PhysRevLett.105.177002, PhysRevLett.105.077001, Alicea2012, Sato2016, 10.21468/SciPostPhysLectNotes.15} and Kitaev spin liquids~\cite{KITAEV20062, Balents2010, Savary_2017, annurev-conmatphys-031218-013401, PhysRevB.103.174417, 10.1093/ptep/ptad115} and so on.
In topological quantum computation, quantum logic gates are implemented by braiding these Majorana quasiparticles, which obey the non-Abelian statistics~\cite{RevModPhys.80.1083, Sarma2015, PhysRevB.95.235305, 10.21468/SciPostPhys.3.3.021, 10.1063/5.0102999}. To implement a braiding operation, positions of the Majorana quasiparticles are controlled by moving the domain walls of the topological number\cite{Alicea2011}. While fast and robust computation requires rapid transferring Majorana quasiparticles with high fidelity~\cite{PhysRevX.3.041017, PhysRevB.91.201102, PhysRevB.91.201404, PhysRevB.91.174305, PhysRevB.99.155150, PhysRevResearch.2.033475, PhysRevApplied.21.034033, yu2025nonadiabaticbraidingmajoranamodes}, fast movement generically reduces fidelity due to non-adiabatic transitions. Thereby, understanding the dynamical properties of edge states by moving domain walls is crucial for developing a superior control method to overcome this speed-fidelity trade off. We note that the transfer of edge states via dynamical motion of domain walls has been demonstrated experimentally in photonic waveguide arrays~\cite{Frank:22}.

In this work, we theoretically propose a high-fidelity protocol to transfer topological edge states by dynamically moving a domain wall between regions with different topological numbers in one dimension.
We focus on a one-dimensional discrete-time quantum walk with two internal states (QW, in short), ~\cite{PhysRevA.48.1687, Meyer1996, Kempe01072003, doi:10.1137/S0097539705447311, Edge_Kitagawa2012} due to its experimental feasibility~\cite{KNIGHT2003147, PhysRevA.68.020301, PhysRevA.69.012310, Roldan15122005, PhysRevA.78.042304, PhysRevLett.108.010502, Crespi2013, PhysRevLett.103.090504, science.1174436, PhysRevLett.104.100503, Matjeschk_2012}. 
The QW exhibits nontrivial topological phases\cite{PhysRevA.82.033429, PhysRevB.84.195139, Kitagawa2012, PhysRevB.88.121406, PhysRevB.102.035418}, whose topological number can be easily tuned in various experiments. This allows us to vary the position of the domain wall by changing parameters over time and space, which makes the edge states to follow the moving domain wall.
We study the dynamics of edge states of QW following the moving domain wall of the topological numbers at a constant velocity. 
This system can be mapped to the uniform linear motion of a particle obeying a Dirac equation\cite{PhysRevA.73.054302, PhysRevA.97.062111}, implying that Lorentz invariance and relativistic effects govern dynamics of edge states.
We propose a protocol for fast and high-fidelity edge states transfer by taking into account these relativistic effects, such as the Lorentz contraction.
Furthermore, we investigate effects of dissipation, such as bit and phase flips of the internal states of the QW, on the transfer efficiency. We reveal an intriguing velocity-dependent robustness of the transfer process in the presence of a specific type of decoherence. This behavior also originates from the relativistic effects of the internal states of the edge states.

This paper is organized as follows. In Sec.~\ref{sec: sec2} we introduce the QW used in this work and show the QW obeys the Dirac equation in the continuum limit. Then, we derive the eigenfunction of the edge state localized near the domain wall of the topological numbers in the static system.  In Sec.~\ref{sec: sec3}, assuming the domain wall moving with a constant velocity, we analytically derive the edge states in both comoving and laboratory frames. In Sec.~\ref{sec: sec4}, we propose a method to realize high-speed and high-fidelity edge-state transport, termed a three-step transfer protocol. Section \ref{sec: sec5} presents numerical results demonstrating the proposed protocol enhances the transfer efficiency of edge states and show the velocity dependence of transfer efficiency in the absence and presence of dissipations. Section \ref{sec: sec6} is devoted to summaries and conclusions.

\section{Definition of QW and edge states}\label{sec: sec2}

We begin by introducing the time evolution of the QW both in the absence and presence of environmental noise. Then, we briefly explain the topological phases of the QW. To explain the emergence of edge states near the topological domain wall, we derive the effective Dirac Hamiltonian obtained from the time-evolution operator of the QW in the continuum limit. Subsequently, the eigenstates of the edge states localized near the domain wall are also analytically derived by solving the Dirac Hamiltonian.

\subsection{\label{sec:level2} Time evolution of QW}

The QW describes the dynamics of quantum states in a Hilbert space
\begin{equation}\label{eq:hilbert}
	\mathcal{H} = \mathcal{H}_{\mathrm{p}} \otimes \mathcal{H}_{\mathrm{in}},
\end{equation}
where $\mathcal{H}_{\mathrm{p}}$ is the Hilbert space spanned by the discrete position basis $\ket{n}$, and $\mathcal{H}_{in}$ is the Hilbert space corresponding to the two internal states spanned by $\ket{L} = (1 \ 0)^{T}$ and $\ket{R} = (0 \ 1)^{T}$. Therefore, the quantum state in $\mathcal{H}$ at discrete time $\tau$  can be expressed as
\begin{equation}\label{eq: state_vec}
	\ket{\psi_\tau} = \sum_{n} \sum_{\sigma = L, R} \psi_{\sigma}(n, \tau) \ket{n} \otimes \ket{\sigma},
\end{equation}
where $\psi_{\sigma}(n, \tau)$ stands for the wavefunction amplitude.
The time evolution of $\ket{\psi_t}$ is governed by a time evolution operator $U$ as
\begin{equation}\label{eq: time_evo}
	\ket{\psi_{\tau +1}} = U\ket{\psi_{\tau}}. 
\end{equation}
The time evolution operator $U$ is defined as
\begin{equation}
 U=S C,
\label{eq: U}
\end{equation}
where $S$ and $C$, referred to as the shift and coin operators, respectively, are defined as
\begin{equation}\label{eq: shift_op}
	S = \sum_{n} \left(\ket{n+1}\bra{n} \otimes \ket{R}\bra{R} + \ket{n-1}\bra{n} \otimes \ket{L}\bra{L}\right),
\end{equation}
\begin{equation}\label{eq: coin_op}
	C = \sum_{n} \ket{n}\bra{n} \otimes 
	\begin{pmatrix}
		\cos \theta & -\sin \theta \\
		\sin \theta & \cos \theta 
	\end{pmatrix}.
\end{equation}
Note that $S$ changes the positions of wavefunctions according to their internal states and $C$ changes only the internal states according to the value of $\theta$. As explained later, $\theta$ depends on the position $n$ and the time $\tau$ in the current work.

Next, we introduce the time evolution of the QW in the presence of dissipation. Let $\rho_{\tau} $ denote the density matrix operator of the QW at time $\tau$. In this work, we assume that the time evolution starts from a pure state $\rho_0= \ket{\psi_0}\bra{\psi_0}$ for simplicity. The time evolution is governed by the following quantum master equation for the density matrix operator,
\begin{equation}\label{eq: evl_diss}
	\rho_{\tau +1} = (1-p) \, U \rho_{\tau} U^{\dag} + p \, K U \rho_{\tau} U^{\dag} K^{\dag},
\end{equation} 
where $p \, \in \, [0,1]$ is a real number that quantifies the strength of dissipation, and $K$ stands for a Kraus operator representing dissipation. Here, we focus on dissipation affecting only the internal states of the QW and introduce three kinds of Kraus operators $K$ defined as
\begin{equation}\label{eq: diss_op}
	K_{j}(\phi) = \sum_{n} \ket{n} \bra{n} \otimes e^{i \phi \sigma_{j}} \, \, \, \,(j = x, y, z),
\end{equation}
where $\sigma_{j=x,y,z}$ are Pauli matrices acting on the internal states and $\phi \in[0,2\pi]$ characterizes the strength of dissipation. Each Kraus operator describes the following dissipation process. The Kraus operator $K_{x}(\phi)$ describes dissipation that alters the superposition of internal states, specifically, $K_{x}(\pm\pi / 2)$ corresponds to a bit-flip dissipation. $K_{z}(\phi)$ describes dephasing process that introduces a phase difference between two internal states. Especially, $K_z{\pm\phi/2}$ corresponds to a phase-flip dissipation. Finally, $K_{y}$ corresponds to the simultaneous effect of $K_{x}(\phi)$ and $K_{z}(\phi)$.

\subsection{Topological phases and edge states of the QW}

The QW is an analogue of a Floquet topological insulator in which non-trivial topological phases are induced by a periodic driving external field\cite{Oka2009,Kitagawa2011}.
In this system,  $\pi$ energy edge states appear in addition to zero-energy edge states. In case of a QW, topological invariants associated with these two types of edge states have been studied in Refs~\cite{PhysRevB.88.121406, Cedzich_2016}. According to Ref~\cite{PhysRevB.88.121406}, the existence of zero- and $\pi$ edge modes is characterized by the invariants $\nu_0$ and $\nu_\pi$, respectively. Their values depend on $\theta$ as
\begin{align}
 (\nu_0, \nu_\pi) = 
\left\{
\begin{array}{ll}
(1,0) & \quad\text{for}\quad  0<\theta<\pi \\
(0,1) & \quad\text{for}\quad -\pi < \theta < 0
\end{array}
\right. .
\end{align}

Since we focus only on the dynamics of the zero energy states, we derive the effective Hamiltonian describing the states near zero energy through the time evolution operator. This approach allows to analytically derive the eigenstate of the zero energy state and, furthermore, unveil Lorentz invariance in this system.

The effective Hamiltonian $H_{\mathrm{eff}}$ of the one-time step time-evolution operator of the QW is derived as the time-independent generator of time evolution operator:
\begin{align}
	U = e^{-i H_{\mathrm{eff}}},
\end{align}
where we set $\hbar=1$.
By taking the continuum limit of $U$ in the position and time spaces\cite{PhysRevA.73.054302}, we can derive $H_{\mathrm{eff}}$ around zero energy. Let $\tilde{U} = \tilde{S} \tilde{C}$ be the continuum limit of $U = SC$, and $\Psi(x, t)$ be a wavefunction of the QW in the continuum limit at the  position $x$ and at the time $t$ ($x,t \in \mathbb{R}$). Since the operator $\tilde{U}$ represents an infinitesimal translation operator in the time direction, $\tilde{U}$ and $\Psi(x, t)$ satisfy
\begin{align}\label{eq: lim of time_evo}
	\Psi(x, t + \delta) = \tilde{U} \Psi(x, t).
\end{align}
where $\delta (\rightarrow 0)$ is an infinitesimally small parameter.

Next, we consider $\tilde{S}$ and $\tilde{C}$, the continuum limits of the shift operator $S$ and coin operator $C$. First, we discuss $\tilde{S}$. Let $D \ (D^{-1})$ be the operator that translates the position of wavefunction to the left (right) infinitesimally. Using these operators, $\tilde{S}$ can be expressed as
\begin{align}\label{eq: inf_shift}
	\tilde{S} = \frac{1}{2} (1 + \sigma_{z}) D + \frac{1}{2} (1 - \sigma_{z}) D^{-1}.
\end{align}
Furthermore, $D$ and $D^{-1}$ satisfy the following relations,

\begin{align}\label{eq: D_eq}
	D\Psi(x, t) = \Psi(x + \delta, t) ,\\
	D^{-1}\Psi(x, t) = \Psi(x - \delta, t),
\end{align}
respectively.
By applying a Taylor expansion to the right-hand side of the above equations with respect to position $x$, $D$ and $D^{-1}$ can be described as

\begin{align}
	D =  1 + i \delta \hat{p}, \label{eq: op_D} \\
	D^{-1} =  1 - i \delta \hat{p} \label{eq: op_D-1},
\end{align}
where $\hat{p}$ is a momentum operator. Substituting Eq.~\eqref{eq: op_D},~\eqref{eq: op_D-1} to Eqs.~\eqref{eq: inf_shift}, 
\begin{align}
	\tilde{S} = e^{i \delta \hat{p} \sigma_{z}}
\end{align}
can be obtained.

Secondly, we consider the continuous coin operator $\tilde{C}$. The discrete coin operator $C(\theta)$ can be expressed using the exponential function as follows:
\begin{align}
	C(\theta) = e^{-i \theta \sigma_{y}}.
\end{align}
Now we assume that $|\theta| \ll 1$. In this case, by applying a Taylor expansion to $C(\theta) = e^{-i [\theta / \delta] \delta \sigma_{y}}$,
\begin{align}
	\tilde{C}(\theta) = 1 - i [\theta / \delta] \delta \sigma_{y}
\end{align}
can be obtained. Therefore, $\tilde{U} = \tilde{S} \tilde{C}(\theta)$ can be written as
\begin{align}\label{eq: colim_U}
	\tilde{U} = 1 + \delta \left[
	 i  \hat{p} \sigma_{z} 
	- i \delta^{-1} \theta  \sigma_{y}
	 \right].
\end{align}
Finally, from Eq.~\eqref{eq: lim of time_evo} and~\eqref{eq: colim_U}, the continuum limit of the time evolution equation can be expressed as
\begin{align}\label{eq: dirac_evo}
	i\frac{\partial}{\partial t}\Psi(x, t) &\equiv H_{\mathrm{eff}} \Psi(x, t) \\
	&= \left[ i \frac{\partial}{\partial x} \sigma_{z}
	  + \delta^{-1} \theta \sigma_{y} \right] \Psi(x, t),
\end{align}
 Note that Eq.~\eqref{eq: dirac_evo} corresponds to the Dirac equation in (1+1) dimensions. We also note that the effective speed of light $\tilde{c}$ is one, since the coefficient of momentum operator in Eq.~\eqref{eq: dirac_evo} is one.

\begin{figure}[t]
	\centering
	\includegraphics[
		width=0.88\columnwidth,
		height=0.35\textheight,
		keepaspectratio
	]{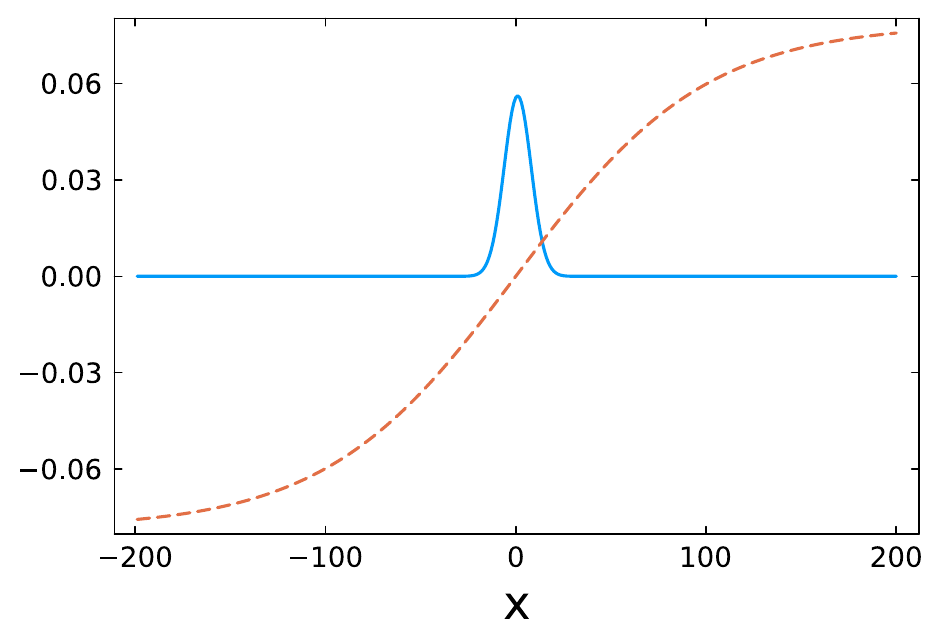}
	\caption{The edge state of the QW (solid curve) with $\theta(x) = \theta_{0} \tanh (ax) / 10$, where $\theta_0 = \pi / 4, a = 0.01$ (dashed curve). The edge state emerges near $\theta(x) \approx 0$.}
	\label{fig:edge_tanh}
\end{figure}

%
As we explained at the beginning of this subsection, topological numbers are changed when $\theta$ changes the sign. 
Therefore, we consider the following position dependent $\theta$,
\begin{align}\label{eq: theta_v0}
	\theta (x) = \theta_{0} \tanh (ax),
\end{align}
to realize the domain wall of the topological numbers appearing around $x = 0$, as illustrated in Fig.~\ref{fig:edge_tanh}. Let $\Psi_{0}(x)$ be the wavefunction of edge state when $\theta$ is defined in Eq.~\eqref{eq: theta_v0}. To analytically derive the eigenstates of the edge state, we consider the linear term of $x$ of $\theta(x)$ in Eq.\ (\ref{eq: theta_v0}) by assuming $ a x \ll 1$. In this case, the effective Hamiltonian $H_{\mathrm{eff}}$ is expressed as
\begin{align}\label{eq: H_zero}
	H_{\mathrm{eff}} = -\hat{p} \sigma_{z} + \theta_{0} a \delta^{-1} x \sigma_{y} 
					 \equiv H_{\mathrm{eff}}^{(0)},
\end{align}
and the eigenfunction for the zero-energy sate $\Psi_{0}(x)$ satisfies
\begin{align}
	H_{\mathrm{eff}}^{(0)} \Psi_{0}(x) = 0.
\end{align}
This eigenequation can be transformed into the eigenequation for the Hamiltonian of the harmonic oscillator by considering squared $H_{\mathrm{eff}}$ (see Appendix. A). As a result, $\Psi_{0}(x)$ can be derived as the Gaussian form
\begin{align}\label{eq: psi_zero}
	\Psi_{0}(x) = \frac{1}{\sqrt{2\xi \sqrt{\pi}}} 
	\exp  \left( -\frac{\it{x}^{\rm{2}}}{2\xi^{2}} \right)
	\begin{pmatrix}
		1 \\
		-1
	\end{pmatrix},
\end{align}
where 
\begin{align}
\xi = \frac{1}{\sqrt{\theta_{0} \it{a} \delta^{-\rm{1}}}},
\end{align}
represents the localization length of the edge state. The edge state in Eq.\ (\ref{eq: psi_zero}), termed the stationary edge state hereafter, is also shown in Fig.\ \ref{fig:edge_tanh}.

\section{Analytical Results of the Wavefunction of moving Edge State}\label{sec: sec3}

Here, we analytically derive the wavefunction of an edge state localized at the moving topological domain wall with a constant velocity, using the effective Hamiltonian of the QW in Eq.\ (\ref{eq: H_zero}) and the zero-energy edge state of the QW in Eq.\ (\ref{eq: psi_zero}). We remark that the Lorentz invariance of the Dirac equation plays an essential role in this derivation~\cite{SHYTOV20091087, PhysRevX.3.041017, oka2025}.

As mentioned in previous section, the edge state appears near the domain wall. Therefore, by introducing a time-dependent parameter $\theta(x,t)$ of the form
\begin{align}\label{eq: tanh_vt}
	\theta(x, t) = \theta_{0} \tanh [a (x - vt)],
\end{align}
and performing the time evolution of the edge state in Eq.\ (\ref{eq: psi_zero}), we may expect that the edge state follows this moving domain wall with the constant velocity $v$. Taking the above $\theta(x,t)$, the effective Hamiltonian is given by
\begin{align}\label{eq:H_D}
	H_{\mathrm{eff}}
    \equiv H_{\mathrm{eff}}^{(v)} (x, t)
	= -\hat{p} \sigma_{z} + \theta_{0} \delta^{-1} 
      a(x - vt) \sigma_{y}.
\end{align}
The time-dependent wavefunction of the edge state $\Psi^{(v)} (x, t)$ satisfies
\begin{align}\label{eq: dirac_lab}
	i \frac{\partial}{\partial t} \Psi^{(v)}(x. t)
	= H_{\mathrm{eff}}^{(v)} (x, t) \Psi^{(v)}(x. t).
\end{align}
While solving $\Psi^{(v)}(x,t)$ analytically is generally non-trivial due to time dependence of the Hamiltonian, the time-dependent Hamiltonian in Eq.\ (\ref{eq:H_D}) can be mapped to the time independent system by the Lorentz invariance because the current system corresponds to uniform linear motion of a particle obeying a Dirac equation. In the comoving frame, i.e., a coordinate moving at a constant velocity $v$ together with the domain wall, the wavefunction can be derived for the time-independent Hamiltonian as explained following. Let $(x^{'}, t^{'})$ and $(x, t)$ represent the position and time coordinates in the comoving and laboratory frames, respectively. They are related by the following Lorentz transformation:
\begin{align}
	x^{\prime} = \gamma (x - vt) \label{eq: Lo_x} \\
    t^{\prime} = \gamma (t - vx) \label{eq: Lo_t},
\end{align}
where the Lorentz factor $\gamma = 1 / \sqrt{1 - v^2}$. Recalling that the effective speed of light $\tilde{c}$ is $\tilde{c}=1$ and applying Eqs.~\eqref{eq: Lo_x} and~\eqref{eq: Lo_t}, we derive the time-independent Hamiltonian $H_{\mathrm{eff}}^{'}(x)$ in the comoving frame,
\begin{align}\label{eq: H_dash}
	H_{\mathrm{eff}}^{'}(x)
	= - \gamma \hat{p} \sigma_{z} 
      + \theta_{0} \delta^{-1} \frac{a}{\gamma} x^{'} \sigma_{y}.
\end{align}
Comparing Eq.~\eqref{eq: H_dash} with Eq.~\eqref{eq: H_zero}, the transformation effectively scales the parameter $a$ by a factor of $1/\gamma$ relative to the static case. Therefore, the wavefunction in the comoving frame $\Psi^{\prime}(x^{\prime})$ is obtained by replacing $\xi$ with $\xi/\sqrt{\gamma}$ in Eq.~\eqref{eq: psi_zero},
\begin{align}\label{eq: psi_dash}
	\Psi^{\prime}(x^{\prime}) 
	= 
	\frac{1} {\sqrt{2 \xi \sqrt{\pi / \gamma}}} {\rm{exp}} 
		            \left( - \frac{ {x^{\prime}}^{2}} {2 (\xi / \sqrt{\gamma})^{2}} \right) 
		            \begin{pmatrix}
		            	1 \\
		            	-1
		            \end{pmatrix}
		            .
\end{align}
We remark that the shrinking of the localization length by the factor $1/\sqrt{\gamma}$ in the comoving frame originates from the Lorentz contraction. 

Finally, the wavefunction of the moving edge state in the laboratory frame, $\Psi^{(v)}(x, t)$, is derived by applying the inverse Lorentz boost to Eq.~\eqref{eq: psi_dash}. Let $S$ be the operator that transforms $\Psi^{\prime} (x^{\prime})$ into $\Psi^{(v)}(x, t)$ under the Lorentz boost:
\begin{align}\label{eq: spnr_trs}
	\Psi^{(v)}(x, t) = S \Psi^{\prime}(x^{\prime}).
\end{align}
For the Lorentz boost in the $x$ direction, $S$ takes the form~\cite{sakurai1967advanced}
\begin{align} \label{eq:nonum}
	S = \exp \left[ -\frac{\eta}{2}  \, \hat{\gamma}^{0} \hat{\gamma}^{1}   \right],
\end{align}
where $\eta= \mathrm{arctanh} \, v$ is the rapidity and $\hat{\gamma}^{i} \ (i = 0, 1)$ are the gamma matrices. Given that the gamma matrices in Eq.~\eqref{eq:nonum} are $\hat{\gamma}^{0} = \sigma_{y}$ and $\hat{\gamma}^{1} = i \sigma_{x}$, $S$ is written down as
\begin{align}\label{eq: S}
	S = \exp \left[ \frac{1}{2} \eta \sigma_{z} \right] = 
	\sqrt{\gamma} \ \mathrm{diag} \left( \sqrt{1 + v}  \ \ \sqrt{1 - v} \right).
\end{align}
Consequently, from Eqs.~\eqref{eq: spnr_trs} and~\eqref{eq: S}, the wavefunction of an edge state moving with velocity $v$ is derived as
\begin{align}\label{eq: Psi_v}
	\Psi^{(v)} (x, t)
	= \frac{\gamma^{3/4}} { (2 \xi \sqrt{\pi})^{1/2} }
	\exp \left[ - \frac{\it{\gamma^{\rm{2}} (x - vt)^{\mathrm{2}}}} {2 (\xi / \sqrt{\gamma})^{2}} \right]
	\begin{pmatrix}
		\sqrt{1 + v} \\
		-\sqrt{1 - v}
	\end{pmatrix}
	.
\end{align}
We call the above edge state the relativistic edge state.
We emphasize that the internal state defined in Eq.~\eqref{eq: Psi_v} depends on $v$, due to the relativistic effect.

\begin{figure}[tbp]
	\centering
	\includegraphics[width=1.0\columnwidth]{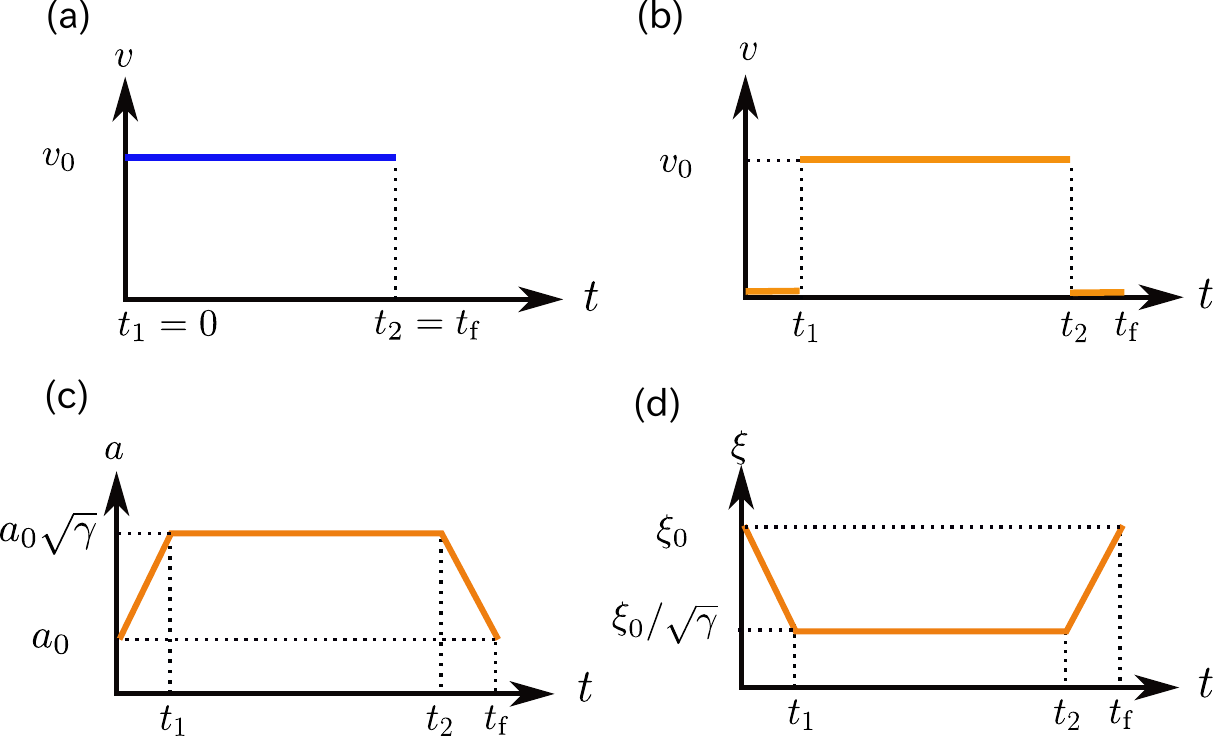}
	\caption{Schematic diagrams of the time evolution of the parameters, the velocity $v$ and the slope of the domain wall $a$ during the edge state transfer. (a) The velocity $v$ in one-step transfer. (b) The velocity $v$ in three-step transfer. (c) The slope $a$ in three-step transfer. (d) The corresponding localization length $\xi$ in three-step transfer.}
	\label{fig: cp13_four}
\end{figure}

\begin{figure*}[tbp]
	\centering
	\includegraphics[width = 1.0\textwidth]{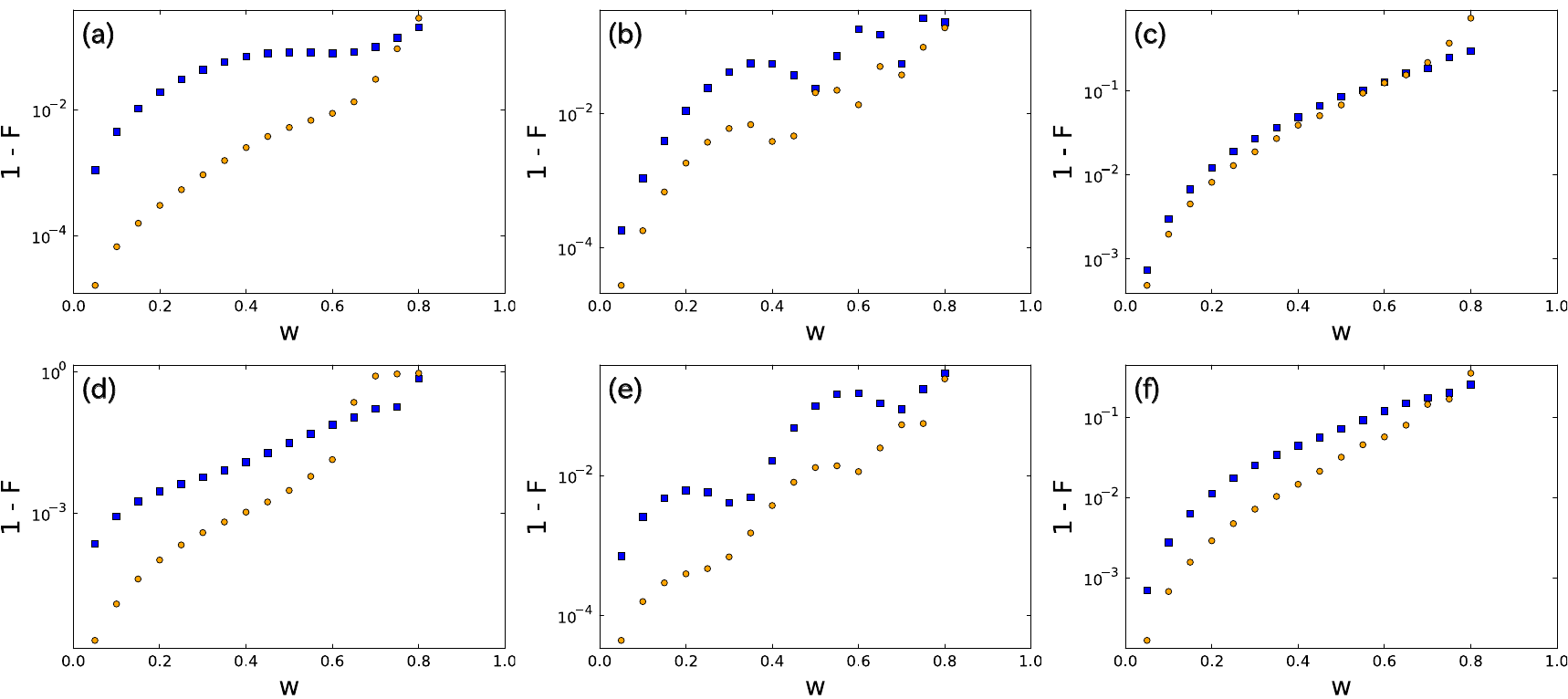}
	\caption{Transfer velocity dependence of the infidelity, $1-F$, for the edge state using one-step and three-step transfer protocols. The horizontal axis represents normalized velocity $\tilde{v} = (t_2 - t_1)v / t_{f}$. For the three-step transfer protocol, we set $t_{1} = 10$, $t_{2} = 110$, and $t_{f} = 120$, while for one-step transfer protocol, $t_{1} = 0$ and $t_{2} = t_{f} = 100$. Squares represent the results for the one-step transfer, and circles represent those for the three-step protocol for (a) $\theta_{0} = \pi / 6, a = 0.01$. (b) $ \theta_0 = \pi/6, a = 0.1$. (c) $\theta_0 = \pi/6, a = 1.0$. (d) $\theta_0 = \pi / 12, a=0.01$. (e) $\theta_0 = \pi / 12, a = 0.1$. (f) $\theta_0 = \pi / 12, a = 1.0$. }
	\label{fig: compare13}
\end{figure*}

\section{Proposal of three-step transfer protocol}\label{sec: sec4}


As discussed in the previous section, the edge state in the QW can be transferred by the time evolution of the initial state defined in Eq.~\eqref{eq: psi_zero} driven by the time-dependent $\theta$ given by Eq.~\eqref{eq: tanh_vt}. In this work, we refer to this transfer protocol as the one-step transfer protocol. Figure \ref{fig: cp13_four}.(a) illustrates the time dependence of the velocity $v$ in this protocol. However, nonadiabatic transitions are inevitable when the stationary edge state is suddenly accelerated. To address this, we propose a protocol termed a three-step transfer that suppresses such transitions to enables high-speed and higher-fidelity transport. Our protocol based on principle to gradually transform the wavefunction of the stationary edge state in Eq.\ (\ref{eq: psi_zero}) into that of the relativistic edge state in Eq.\ (\ref{eq: Psi_v}) before spatially moving the domain wall. This protocol consists of three distinct steps: the first step (from $\tau = 0$ to 
 $\tau = \tau_1$), the second one (from $\tau = \tau_1$ to $\tau = \tau_2$), and the third one (from $\tau = \tau_2$ to $\tau = \tau_{\mathrm{f}}$), as illustrated in Fig.~\ref{fig: cp13_four}.(b)-(d). Each step performs the following operation to the edge state:
\begin{itemize}
 \item Step 1 $(0\le \tau < \tau_1)$:  The stationary edge state at the initial position is adiabatically transformed into the wavefunction form of the relativistic edge state by changing the slop $a$ but keeping its position.
 \item Step 2 $(\tau_1 \le \tau < \tau_2)$: The edge state is transferred by moving the domain wall at a constant velocity $v$ as in the one-step transfer protocol, but starting from the mapped to the relativistic edge state.
 \item Step 3 $(\tau_2 \le \tau_{\mathrm{f}})$: The relativistic edge state is adiabatically transformed into the stationary edge state at the position $v(\tau_2-\tau_2)$, as the inverse transformation of the Step 1.
\end{itemize}

We explain the details of the above operations.
By comparing $\Psi_{0}(x)$ in Eq.~(\ref{eq: psi_zero}) and $\Psi^{(v)}(x, t)$ in Eq.~(\ref{eq: Psi_v}), difference between the stationary and relativistic edge sates are i) the localization length $\xi$ for $\Psi_0$ and $\xi/\sqrt{\gamma}$ for $\Psi^{(v)(x,t)}$ due to Lorentz contraction, and ii) the internal states $(1,1)^T$ for $\Psi_0$ and $(\sqrt{1+v},\sqrt{1-v})^T$ for $\Psi^{(v)}(x,t)$.
The difference in the localization length can be compensated by adiabatically changing the slope parameter $a$ of $\theta(x,t)$ in Eq.~({\ref{eq: tanh_vt}}) at the Step 1 as
\begin{align}
	a(\tau) = a_{0} + a_{0} (\gamma - 1) \frac{\tau}{\tau_1}, \ \ \ (\tau \in [0, \tau_1]),
\label{eq:a_tau}
\end{align}
where $a = a_0$ denotes the initial slope, which correspond to the edge state in the static case. A sufficiently long time $\tau_1$ is required to ensure the adiabatic transformation. Additionally, the internal state is changed by applying the following coin operator to the initial state of $\Psi_0(x)$:
\begin{align}
	C_{v} =  \frac{1}{2}
	\begin{pmatrix}
		\sqrt{1 + v} + \sqrt{1 - v}  &  -\sqrt{1 + v} + \sqrt{1 - v} \\
		 \sqrt{1 + v} - \sqrt{1 - v}  &  \sqrt{1 + v} + \sqrt{1 - v} 
	\end{pmatrix}
	.
\label{eq:C_v}
\end{align}
This operator acts the internal states that aligns the static ones with the relativistic ones.
At the Step 3, the inverse of the transformations in Eqs.~(\ref{eq:a_tau}) and (\ref{eq:C_v}) are applied to the relativistic state at $t_2$ to stop at the postion $v(\tau_2-\tau_1)$.

\section{Numerical Results}\label{sec: sec5}

In this section, we numerically evaluate the efficiency of our proposed protocol for edge-state transport. To this end, we first define fidelity as a measure of the transfer efficiency. Then, we present the velocity dependence of the fidelity in both the absence and presence of dissipation.

Consider the case where the edge state is transferred over a distance of $v (t_2-t_1)$. Note that $t_1=0$ in case of the one-step transfer. In this work, fidelity, or transfer efficiency, is defined as follows~\cite{UHLMANN1976273}
\begin{align}
	F = \mathrm{Tr} \left[ 
	\sqrt{ \rho_{\mathrm{f}}(x, t_{\mathrm{f}}) }
	\sqrt{ \rho_{0} (x - v(t_2-t_1))}
	\right],
\end{align}
where ${ \rho_{\mathrm{f}}(x, t_{f})}$ is the density operator of the state at $t_f$ and ${ \rho_{0} (x - v(t_2-t_1))}$ is the density operator of the edge state at $t=0$ with additional shift in a position space $v (t_2-t_1)$. Fidelity $F$ takes a value between zero and one, with a value closer to one indicating higher transfer efficiency of the edge state. In particular, when we consider only the pure states, $F$ can be expressed in term of the state vector,
\begin{equation}\label{eq: fid_vec}
    F = \left| \Braket{\psi_{f}(x, t_{f}) | \Psi_{0}(x - v(t_2-t_1))} \right|.
\end{equation} 
\subsection{Transfer without dissipation}
\begin{figure}[tbp]
	\centering
	\includegraphics[width = 0.8\columnwidth]{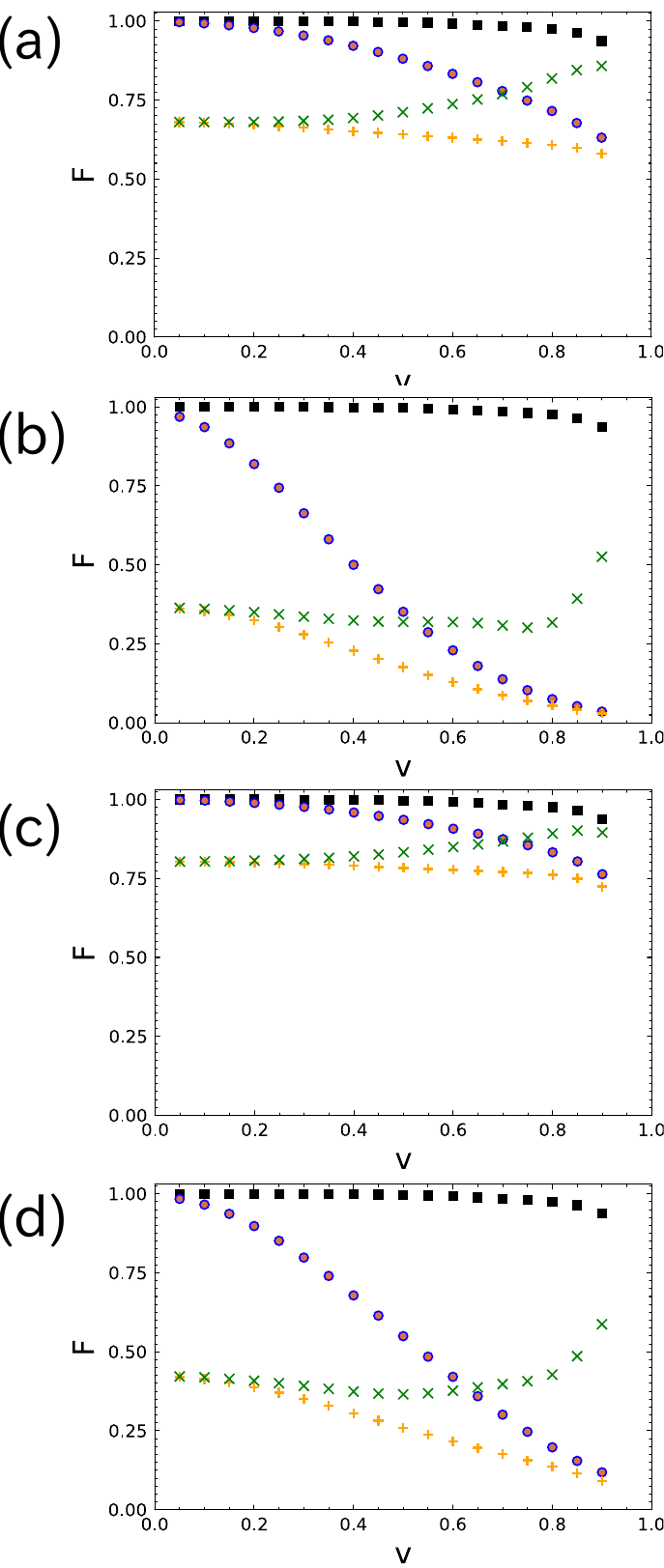}
	\caption{Transfer-velocity dependence of the infidelity, $1-F$, when the edge state is transferred with dissipation using the three-step transfer protocol with $\theta_{0} = \pi / 4$, $a = 0.01$, $t_1 = 10$, $t_2 = 110$, and $t_{f} = 120$ for (a) $\phi = \pi / 2$ and $p = 0.01$. (b) $\phi = \pi / 2$ and $p = 0.1$. (c) $\phi = \pi / 4$ and $p = 0.01$. (d) $\phi = \pi / 4$ and $p = 0.1$. Black Squares correspond to the results in the absence of dissipation, magenta circles correspond to bit flip dissipation, green x-shaped markers indicate dephasing dissipation, and yellow plus-shaped markers represent the case where both bit flip and dephasing dissipation affect simultaneously.}
	\label{fig: dissipation}
\end{figure}

First, we compare the fidelity of the edge state transferred by the three-step and the one-step transfer protocols without dissipation. Figure \ref{fig: compare13} shows the infidelity $1-F$ as a function of the normalized velocity $\tilde{v}=v(t_2-t_1)/t_f$. Note that $\tilde{v}=v$ for the one-step protocol, while $\tilde{v}<v$ for the three-step protocols. Therefore, the normalized velocity accounts for the difference effective transport durations between the two protocols, allowing for a fair comparison of their efficiency.

Figure \ref{fig: compare13} shows that the three-step protocol remarkably increases the fidelity of edge state transfer for broad normalized velocity space with various parameter sets. In particular, at moderate velocities $\tilde{v}$, the fidelity significantly increases by a factor $10^2$ to $10^3$ to the one-step protocol. Furthermore, at the smaller $\tilde{v}$, the fidelity becomes larger for the smaller values $a$ and $\theta_0$. This can be explained by the facts that the lower velocity reduces non-adiabatic transition and smaller $a$ and $\theta_0$ ensure that the relativistic edge state is well approximated by the solution in the continuum limit in Eq.\ (\ref{eq: Psi_v}). 

\subsection{Transfer with dissipation}
Figure.~\ref{fig: dissipation} compares the transfer efficiency of the edge states by three-step transfer as a function of the normalized velocity with dissipation using Eqs.~\eqref{eq: evl_diss} with that without dissipation. We consider three kinds of the Kraus operators, $k_x(\phi)$, $k_y(\phi)$, and $K_z(\phi)$ as well as several parameter sets of the strength of dissipation$\phi=\pi/2, \pi/4$ and $p=0.01, 0.1$. 

First, comparing the results with $p=0.01$ and $0.1$, we observe that the fidelity is reduced as increasing the strength of dissipation $p$.
In addition, the fidelity also decreases when the dissipation strength $\phi$ approaches $\pi/2$. 
However, remarkably, bit-flip noise $(K_x)$ has a negligible effect on the fidelity in the low-velocity regime, irrespective of the dissipation strength of $p$ and $\phi$.
Furthermore, dephasing noise $(K_z)$ becomes suppressed as $v$ approaches the effective speed of light.

These behaviors can be understood by considering the velocity-dependence of the internal states of the relativistic edge state in Eq.\ (\ref{eq: Psi_v}).
In the laboratory frame, in the limit of $\tilde{v}\rightarrow 0$, the internal state reduces to
\begin{align}
	\begin{pmatrix}
		\sqrt{1 + v} \\
		-\sqrt{1 - v}
	\end{pmatrix}
	\rightarrow
	\begin{pmatrix}
		1 \\
		-1
	\end{pmatrix}
	,
\end{align}
which is an eigenstate of $\sigma_x$ whose eigenvalue is $-1$.
Consequently, the state is inherently robust against $\sigma_x$-type bit-flip errors at low velocities. 
We emphasize that this property is ensured by chiral symmetry in the context of topological insulators: edge states are also the eigenstates of the unitary operator $\Gamma$, which satisfies $\Gamma H \Gamma^{-1} = -H$, representing chiral symmetry. 
Similarly, in the high-velocity limit $v \rightarrow 1$, the internal state of the moving edge state becomes
\begin{align}
	\begin{pmatrix}
		\sqrt{1 + v} \\
		-\sqrt{1 - v}
	\end{pmatrix}
	\rightarrow \sqrt{2} 
	\begin{pmatrix}
		1 \\
		0
	\end{pmatrix}
	,
\end{align}
which is an eigenstate of $\sigma_{z}$, whose eigenvalue is one. 
In this high-velocity regime, the wavefunction becomes an eigenstate of the dephasing operator $\sigma_z$,
meaning that the edge state is robust against $\sigma_z$-type dephasing errors.
Note that this property is ensured by Lorentz invariance of the Dirac equation.

\section{summary}\label{sec: sec6}
In this work, we theoretically proposed a high-fidelity protocol for transferring topological edge states by dynamically moving the topological domain wall in the one dimension. 
Taking the advantage of experimental feasibility, we focused on the one-dimensional QW with two internal degrees of freedom and investigate the dynamical properties of the edge states.

By considering the continuum limit and Lorentz invariance, we analytically derived the wavefunction of the edge states localized at the moving domain wall with the constant velocity. 
Based on these results, we proposed the three-step transfer protocol which can reduce the non-adiabatic excitations to other states by taking Lorentz contraction of the localization length of the edge states into account.

Our numerical calculation verified the substantial improvement of the efficiency of the three-step transfer protocol by comparing with the naive one-step transfer protocol.
Furthermore, we investigated the velocity dependence of transfer efficiency under dissipation. We found that a bit flip dissipation is suppressed in the low-velocity limit, while the effects of dephasing are reduced as increasing $v$ up to $v=1$. These behaviors can be explained by the velocity-dependent internal state of the edge state, which originate from relativistic effects. 

In the present work, as a first step, we propose the high-speed and high-fidelity transfer protocol of a single edge state. If our protocol is applied for the quantum state transfer of qubit information, the protocol is needed to extend to transfer two states. We will keep it to the next work. While the current work focuses on the high-speed transport of a single edge state, extending this protocol to multiple states or the transfer of entangled qubit information remains an important next step.

\acknowledgments

We thank Y.\ Asano, T.\ Sano, and A.\ Sasaki for their helpful discussions.
This work was supported by JSPS KAKENHI (Grants Nos. JP23K22411, JP24K00545, and JP26K00624).
H.O. was supported by JST PRESTO Grant No. JPMJPR2454. N.K. was supported by the RIKEN TRIP initiative.

\appendix

\section{\texorpdfstring{Derivation of Eq.~\eqref{eq: psi_zero}}{Derivation of Eq. psi zero}}

In this appendix, we derive Eq.~\eqref{eq: psi_zero}. The calculation method is based on Ref.\ \onlinecite{PhysRevX.3.041017} that discusses a unitary-equivalent Hamiltonian in relation to the effective Hamiltonian of the QW.

Since $\theta$ is time-independent, we consider the time-independent Schr\"{o}dinger equation with eigenenergy $E_n$
\begin{align}
	H_{\mathrm{eff}} \Psi_n(x) = E_n \Psi_n(x).
\end{align}
Squaring $H_{\mathrm{eff}}$ becomes
\begin{align}\label{eq: Heff_sqrd}
	H_{\mathrm{eff}}^2 &= \hat{p}^2 + (\theta_{0} a \delta^{-1})^2 \hat{x}^2
	                     + \theta_{0} a \delta^{-1} \sigma_x. \\
	                   &= H_{0}^2 + \theta_{0} a \delta^{-1} \sigma_x,
\end{align}
where $H_0 = \hat{p}^2 + (\theta_{0} a \delta^{-1})^2 \hat{x}^2$ corresponds to the Hamiltonian for a harmonic oscillator potential. Furthermore, for simplicity, we apply the unitary rotation to $H_{\mathrm{eff}}^2$ defined by $X = e^{i \pi \sigma_{y} / 4}$. This yields
\begin{align}\label{eq: Hefftld}
	\tilde{H}_{\mathrm{eff}}^2 &= X H_{\mathrm{eff}}^2 X^{\dag} \\
	&=  \tilde{H}_{0}^2 + \theta_{0} a \delta^{-1} \sigma_z,
\end{align}
where $\tilde{H}_{0}^2 = X H_{0}^2 X^{\dag} = H_0^2$.
Let $\tilde{\Psi}_n(x) = X \Psi_n(x) = \left(\psi_n(x), \varphi(x)_n \right)^T$ be the two-component eigenfunction of $\tilde{H}_{\mathrm{eff}}$, which satisfies the following eigenvalue equation
\begin{align}\label{eq: egneq_Psitld}
	\tilde{H}_{\mathrm{eff}}^2 \tilde{\Psi}_n(x) = E_{n}^2 \tilde{\Psi}_n(x).
\end{align}
Since $\tilde{H}_{\rm{eff}}^2$ has a diagonal form, the above equation can be rewritten as the simultaneous equations with the same eigenvalue $E_n^2$,
\begin{equation}\label{eq: smleq_tldPsi}
		\begin{aligned}
			& \tilde{H}_{0}^2 \psi_n(x) = ( E_{n}^2 - \theta_{0} a \delta^{-1}) \psi_n(x),  \\
			& \tilde{H}_{0}^2 \varphi_n(x) = ( E_{n}^2 + \theta_{0} a \delta^{-1}) \varphi_n(x) .
		\end{aligned}	
\end{equation}
Since $\tilde{H}_0^2$ is the harmonic oscillator Hamiltonian, $\psi(x)$ and $\varphi(x)$ are given by one of the eigenfunctions of the harmonic oscillator.
Here, we write down $\tilde{\Psi}(x)$ as
\begin{align}\label{eq: Psi_harm}
	\tilde{\Psi}_n(x) = 
	\begin{pmatrix}
		\phi_{l}(x) \\
		\phi_{m}(x)
	\end{pmatrix},
\end{align}
where $\phi_l(x)$ is the $l$-th eigenfunction of $\tilde{H}_{0}^2$,
\begin{align}
		\phi_{l}(x) = C_{l} \mathcal{H}_{l} 
	\exp \left[ - \frac{x^2}{2\xi^2}\right],
\end{align}
whose eigenvalue is 
\begin{align}
	\tilde{E}_{l}^{2} =  \omega \left( l + \frac{1}{2} \right),
\end{align}
where $\omega=2 \theta_0 a \delta^{-1}$, $C_l$ is the normaization constant, and $\mathcal{H}_{n}$ is the Hermite polynomial.

By substituting Eq.~\eqref{eq: Psi_harm} into Eq.~\eqref{eq: egneq_Psitld}, we obtain
\begin{equation*}
		\begin{aligned}
			& \omega\left(l+\frac{1}{2}\right) \phi_l(x) = \left( E_{n}^2 - \frac{\omega}{2}\right) \phi_l(x),  \\
			& \omega\left(m+\frac{1}{2}\right) \phi_m(x) = \left( E_{n}^2 + \frac{\omega}{2}\right) \phi_m(x).
		\end{aligned}	
\end{equation*}
Subtracting these equations yealds $m-l=1$. To ensure that the energy $E_n^2$ is non-negative, we set $l=n-1$ and $m=n$ for integer $n\ge 0$. 
Thus, $\tilde{\Psi}_n(x)$ can be written as
\begin{align}\label{eq: tldPsi_mdfed}
		\tilde{\Psi}(x) = 
	\begin{pmatrix}
		\phi_{n-1}(x) \\
		\phi_{n}(x)
	\end{pmatrix}
	.
\end{align}
%
%
Note that the upper component vanishes if $n=0$.
Substituting Eq.~\eqref{eq: tldPsi_mdfed} to Eq.~\eqref{eq: egneq_Psitld}, the eigenenergy $E_{n}$ can be obtained as
\begin{align}
	E_{n} = \sqrt{\omega |n|}.
\end{align}
This indicates that the eigenfunction of the zero-energy edge state is given by $n=0$, that is, 
\begin{align}
	\tilde{\Psi}_{0}(x) = C_{0} \exp \left[ -\frac{x^2}{2\xi^2} \right]
	\begin{pmatrix}
		0 \\
		1
	\end{pmatrix}
	.
\end{align}
Finally, we obtain the zero-energy state of $H_{\rm{eff}}$ by applying the inverse unitary transformation:
\begin{align}
	\Psi_{0}(x) &= X^\dagger \tilde{\Psi}_{0}(x) \\
	&  = \frac{1}{\sqrt{2\xi \sqrt{\pi}}} 
	\exp  \left( -\frac{\it{x}^{\rm{2}}}{2\xi^{2}} \right)
	\begin{pmatrix}
		1 \\
		-1
	\end{pmatrix}.
\end{align}

\bibliographystyle{apsrev4-2}
\bibliography{transfer_v5}
\end{document}